\documentclass[referee]{aa}

\usepackage{amsmath}
\usepackage{amssymb}
\usepackage{latexsym}
\usepackage{subfig,graphicx}
\usepackage[varg]{txfonts}
\usepackage{natbib}
\bibpunct{(}{)}{;}{a}{}{,} 

\newcommand{\kms}{km\,s$^{-1}$}

\titlerunning{Transverse waves observed with Hi-C}
\authorrunning{Morton and McLaughlin}

\begin{document}
\title{Hi-C \& AIA observations of transverse MHD waves in active regions}

\author{R. J. Morton and J. A. McLaughlin} 

\institute{Department of Mathematics \& Information Sciences, Northumbria University, Newcastle Upon Tyne,
NE1 8ST, UK}

\abstract{The recent launch of the High resolution Coronal imager (Hi-C) provided a unique opportunity to study 
the EUV corona with unprecedented spatial resolution. We utilize these observations to investigate the 
properties of low frequency ($50-200$~s) active region transverse waves, whose omnipresence had been 
suggested previously. {The five-fold improvement in spatial resolution over SDO/AIA reveals coronal loops with widths $150-310$~km and that these loops support transverse waves with displacement amplitudes $<50$~km}. However, the results suggest that wave activity in the coronal loops is of low energy, with typical velocity amplitudes $<3$~\kms. An extended time-series of SDO data suggest low-energy wave behaviour is typical of the coronal structures both before and after the Hi-C observations.}


\date{Received /Accepted}

\maketitle

\section{Introduction}
There have now been numerous reports of ubiquitous magneto-hydrodynamic (MHD) waves in the solar 
atmosphere. In particular the periodic, transverse displacement of magnetic flux tubes in both the chromosphere
(\citealp{DEPetal2007}; \citealp{KURetal2012}; \citealp{MORetal2012c}) and in the solar corona 
(\citealp{TOMetal2007}; \citealp{ERDTAR2008}; \citealp{MCIetal2011}).  The chromospheric 
motions are somewhat better resolved as they are observed with high-resolution (<0''.07 per pixel) imagers, 
e.g., space- (Hinode) and ground-based (ROSA, CRISP). The observations of coronal transverse 
waves, however, are typically restricted because of larger spatial resolutions, >0''.5 per pixel, as well as the 
coronal plasma being optically thin. Both these effects contribute to problems with line-of-sight (LOS) integration 
meaning several coronal structures may contribute to emission within a single pixel. This restriction has led to 
some differing observational results on the properties of the transverse waves. For example, velocity amplitudes 
of these waves are reported as $\sim$0.4~km\,s$^{-1}$ in off-limb, 
active region loops (Coronal Multi-Channel Polarimeter/CoMP - \citealp{TOMetal2007}), while observations with 
the Solar Dynamic Observatory (SDO) suggest typical amplitudes of $\sim$5~km\,s$^{-1}$ 
(\citealp{MCIetal2011}). \cite{DEMPAS2012} demonstrated LOS integration of multiple unresolved structures 
would lead to an underestimate of velocity amplitudes. However, their results suggest this does not 
account for the observed differences. Another solution to the problem is that current resolutions 
and cadences are masking the true nature of transverse waves. The data taken with the High resolution Coronal 
Imager (\citealp{CIRetal2013}) provides a unique opportunity to study active region transverse waves at 
high resolution.  

\begin{figure*}[!tp]
\centering
\includegraphics[scale=0.75, clip=true, viewport=1.0cm 1.0cm 21.5cm 10.6cm]{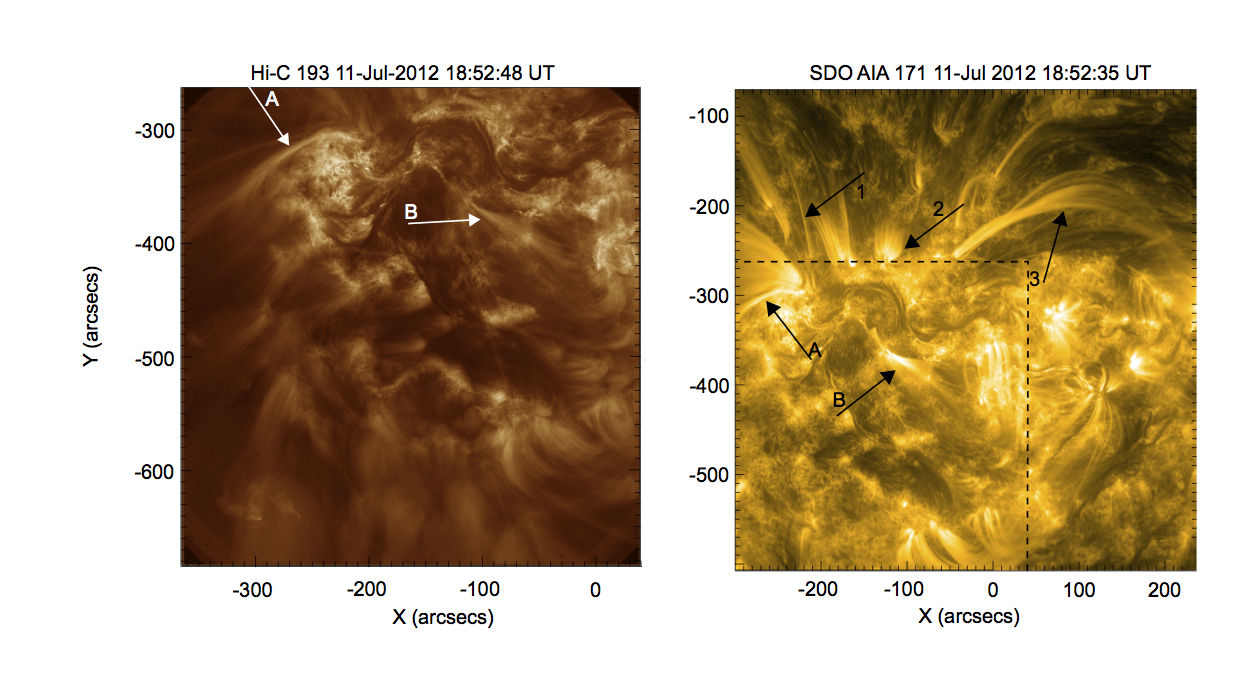} 
\caption{Active region observed with Hi-C $193$~{\AA} (\textit{left}) and SDO/AIA $171$~{\AA} (\textit{right}). 
The displayed AIA field of view demonstrates the extended structure of the active region and the dashed black 
box highlights the overlap with Hi-C. The arrows on 
the Hi-C image highlight the two coronal structures (labelled A and B) studied in high resolution in this paper. The 
same arrows are also shown on the AIA image, with the additional 
arrows labelled 1-3 highlighting the loop structures studied in AIA data only.}\label{fig:fov}
\end{figure*}

\section{OBSERVATIONS AND DATA REDUCTION}
The Hi-C observations took place on 11 July 2012 from {18:51:52}~UT to {18:55:13}~UT centred on an active 
region at (-130''.0,-453''.3) from disk centre\footnote[1]{This is the corrected date for the Hi-C observations. The published paper has a typographical error.}. The images were obtained in the $193$~{\AA} passband (spectral 
width - $5$~{\AA}). The data was taken with a cadence of $\sim5.4$~s and 
a spatial resolution of $0''.103$ per pixel ($\sim$75~km). Further details on the telescope can be found in 
\cite{CIRetal2013}. We obtained the level 1.5, 4k$\times$4k data set through the Virtual Solar Observatory. The 
data was dark-subtracted, flat-fielded, cropped, dust hidden and co-aligned by the Hi-C Science team.  We found 
that the data still displayed visible shifts from frame to frame so we additionally aligned the data using 
cross-correlation to achieve sub-pixel alignment. Testing the accuracy of the alignment suggests the remaining 
errors have a frame to frame RMS value of $0.05$ pixels. The data is missing a frame between 
{18:54:12-18:54:23}~UT so we use linear interpolation to create the missing frame and 
provide a constant sampling rate for wave studies. {Note that the first seven frames of the Hi-C data are inadequate for data analysis. They are displayed in figures and movies but not used for analysis.}

The 193~{\AA} line has strong contributions from Fe XII, which has a peak formation temperature close to 
$1.5$~MK so is ideal for observing coronal features. However, the images not only show well defined coronal 
loops (Fig.~\ref{fig:fov}) but also spicular and fibrillar structures in and around the moss regions. The data shows that the spicules/fibrils are constantly in motion, displaying evidence for flows 
and waves. However, this is contrast to a corona that shows relatively muted dynamic behaviour. To analyse the 
motions of the fine-structure we have to perform some manipulation of the images. Due to the low signal to noise  
(S/N) of Hi-C, we first applied a spatial filtering algorithm to each frame to suppress the highest frequency 
spatial components to increase S/N. We then apply a 5 by 5 box car smoothing function to further suppress 
noise, while retaining sufficient signal to resolve fine-scale structure. To highlight the fine-structure an unsharp 
mask procedure is used (see, e.g, Fig.~\ref{fig:loop}).

Due to the short duration of the Hi-C time-series, we use additional data from the SDO Atmospheric Imaging 
Assembly (AIA) (\citealp{LEMetal2011}) to study the long term behaviour of the region. The data is of a larger region around the Hi-C field of view and comprises of the $171$~{\AA} and $193$~{\AA} bandpasses and covers the period 18:40:47-19:04:47~UT. The AIA data was prepared using the standard techniques, alignment and unsharp masking was also performed on the data sets.


\section{DATA ANALYSIS}
We search the data for signatures of transverse wave motion, i.e., the physical displacement of the structure's 
central axis. This is performed by placing a cross-cut perpendicular to the structure under consideration and 
producing a time-distance diagram. To analyse the structures in the time-distance diagrams, we fit a 
Gaussian function to the cross-sectional flux profile in each time slice (see, e.g., \citealp{ASCSCH2011}), which  
allows for sub-pixel accuracy on locating the central position of the structure's cross-section. {The fitting of the Gaussian requires an estimate of the data noise ($\sigma_N$), which we calculate following \cite{YUANAK2012},
\begin{equation}
\sigma_N=\sqrt{\sigma_p(F)^2+\sigma_d^2+\sigma_r^2+\sigma_{sd}^2}=\sqrt{0.23F+588.4}.
\end{equation}
where $\sigma_{p}(F)$ is the uncertainty in photon noise, $\sigma_d$ dark current, $\sigma_r$ readout and
$\sigma_{sd}$ digitisation (Private communication - A. Winebarger). Similar formulae are used for the AIA 
data noise (\citealp{YUANAK2012}).} We also take into account errors in alignment between the frames. To 
determine the total error on position we add the alignment error to the Gaussian centroid position error, assuming the 
alignment error for each data point to be the RMS value of $0.05$ pixels for Hi-C and 0.01 for AIA data. We 
then supply a non-linear fitting algorithm with the central positions and the standard errors and instruct it to fit a 
function of the form
$$
F(t)=A\sin(\frac{2\pi}{P}t-\phi)+g(t),
$$ 
where $A, P$ and $\phi$ are the displacement amplitude, period and phase, respectively, of the wave. The parameter $g(t)$ is a 
linear function that represents any transverse drift of the structure with time, which could possibly be attributed 
to long period transverse waves. From the fitted parameters we calculate the velocity amplitude using the relation 
$v=2\pi A/P$.

\begin{figure*}[!tp]
\centering
\includegraphics[scale=0.78, clip=true, viewport=0.0cm 0.0cm 21.cm 5.0cm]{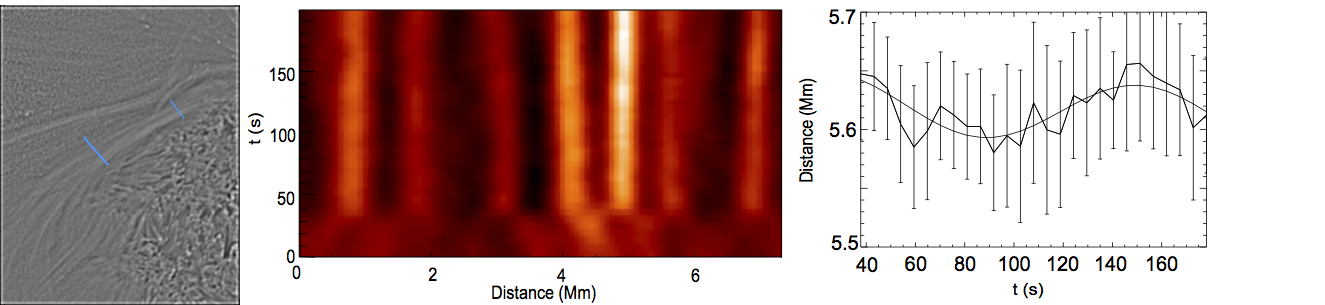} 
\includegraphics[scale=0.78, clip=true, viewport=0.0cm 0.0cm 21.cm 5.0cm]{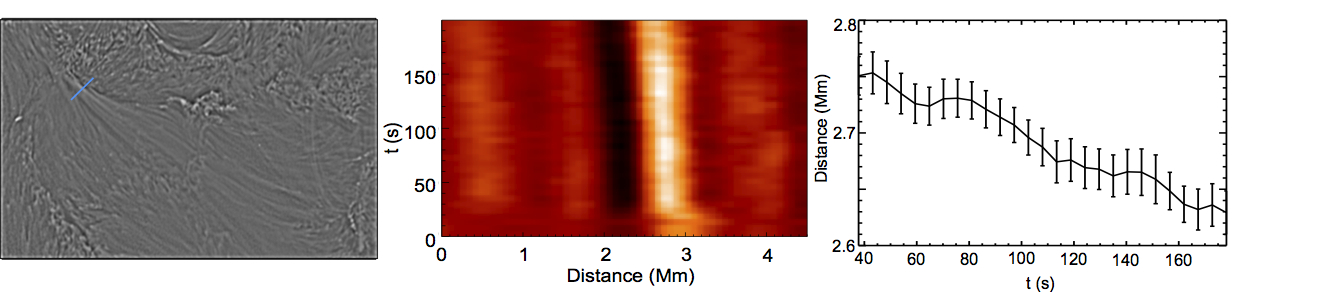} 
\caption{Top row: The \textit{left} panel displays close-up, filtered images of the coronal structures labelled A and the 
\textit{middle} panel displays a time-distance diagram for a cross-cut at the loop apex, with the cross-cut 
position shown in the left panel. The \textit{right} panel is the results of the fitting of a loop at the apex and 
has measured parameters of $P=126\pm80$~s, $A=25\pm22$~km and $v=1.2\pm1.3$~km/s. Bottom row: The \textit{left} panel displays close-up, filtered images of the coronal structures labelled B and the 
\textit{middle} panel displays a time-distance diagram taken close to the footpoint, with the cross-cut position shown in the left panel. The \textit{right} panel is the results of the fitting to the loop.
The time-scales are given in seconds from the start of Hi-C observations.
}\label{fig:loop}
\end{figure*}


Numerous coronal structures can be observed in the Hi-C images. Two large, distinct coronal structures, 
arbitrarily labelled A and B, are easily identified at (-250'',-330'') and (-80'',-370''), respectively (Fig.~\ref{fig:fov} 
- highlighted by arrows). The rest of the coronal structures are fine loops with diffuse emission. {The low 
S/N of the Hi-C data} means any fine-structure of these diffuse loops cannot be resolved. Hence, we concentrate 
on the distinct coronal structures. 

Example time-distance diagrams for A and B are plotted in Fig.~\ref{fig:loop}. 
Perhaps surprisingly, these structures show little visible evidence of periodic transverse motion. For the 
structure A, we show a time-distance diagram taken close to the loop apex (Fig.~\ref{fig:loop} - top panels). The filtering technique has revealed 
this coronal structure consists of 6-7 fine threads. {Fitting a Gaussian to the loop cross-sections gives loop 
widths of 150-310~km, suggesting the fine structure is not well resolved in the corresponding AIA 
$193$~{\AA} images. This is confirmed as the unsharp masking of the AIA images does not reveal these 
individual loop threads (see online movies 1-4). }

\begin{figure}[!tp]
\centering
\includegraphics[width=8.7cm, height=5.3cm, clip=true, viewport=0.0cm 0.0cm 12.cm 8.6cm]{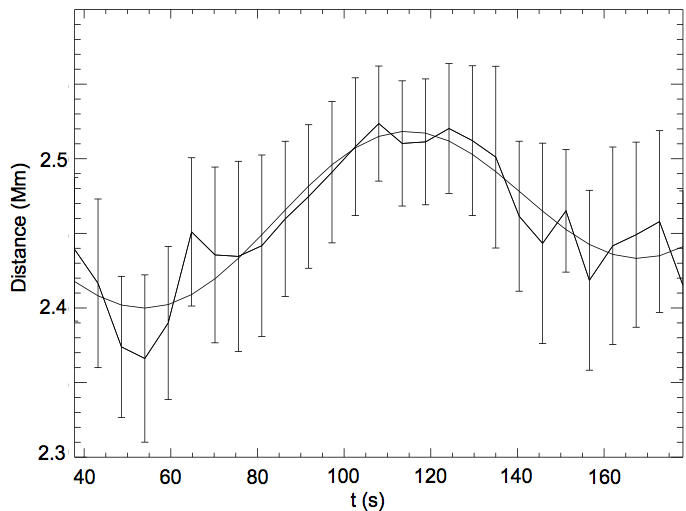} 
\caption{Results from the fitting at the footpoint of structure A (position of cross-cut given in top left panel - Fig.~\ref{fig:loop}). A sinusoidal fit to the data points has measured parameters of $P=109\pm16$~s, $A=50\pm14$~km and $v=2.9\pm0.9$~km/s. 
}\label{fig:oscill}
\end{figure}

{Upon applying the Gaussian fitting technique to the loops we are able to resolve small amplitude 
transverse displacement towards the foot points of the loops in structure A. A sinusiodal fit suggests the displacement is 
periodic (see, e.g., Fig.~\ref{fig:oscill}), where the measured displacement amplitude (47$\pm$14~km) is a factor of 10 smaller than 
the AIA pixel size. The transverse displacements can be measured along a portion of the loop leg 
($\sim2200$~km) but overlapping loops 
prevent us following the signal to the apex. The amplitude of the wave is seen to increase with height along the 
leg, from 25~km to 50~km. Cross-correlation of the signals in neighbouring slits suggests the feature is 
propagating with phase speeds $400\pm300$~\kms. Transverse motion in a neighbouring loop \textbf{in the legs} also 
has a small displacement amplitude of $A=22\pm12$~km, with $P=65\pm10$~s. In contrast, measurements 
suggest the amplitude decreases with height. The identification of periodic transverse displacement towards the 
loop apex is more uncertain due to lower S/N. While the results of the fitting suggest transverse displacements 
with amplitudes of 10-25~km, there are large uncertainties in the fit parameters. The best example of the fits 
towards the loop apex is given in Fig.~\ref{fig:loop}. }

\textbf{The changes in amplitude along the loop legs are interesting.} Cross-correlation suggests the waves are 
propagating, implying an initial increase in amplitude could be due to decreasing density with height. We 
speculate that the reduction in amplitude with height could be explained by spatial damping, e.g., by invoking 
resonant absorption as a damping mechanism for the transverse waves (\citealp{TERetal2010c}; 
\citealp{PASetal2012}). \cite{VERTHetal2010} show resonant absorption is compatible with the decrease in 
propagating wave amplitude along coronal loops observed in CoMP (\citealp{TOMMCI2009}). \textbf{For propagating waves, competition between amplification and damping is expected to occur in 
the majority of loops, with the observable variation in amplitude dependent upon the individual loop.}

\begin{figure}[!tp]
\centering
\includegraphics[scale=0.45, clip=true, viewport=0.0cm 0.0cm 17cm 20.cm]{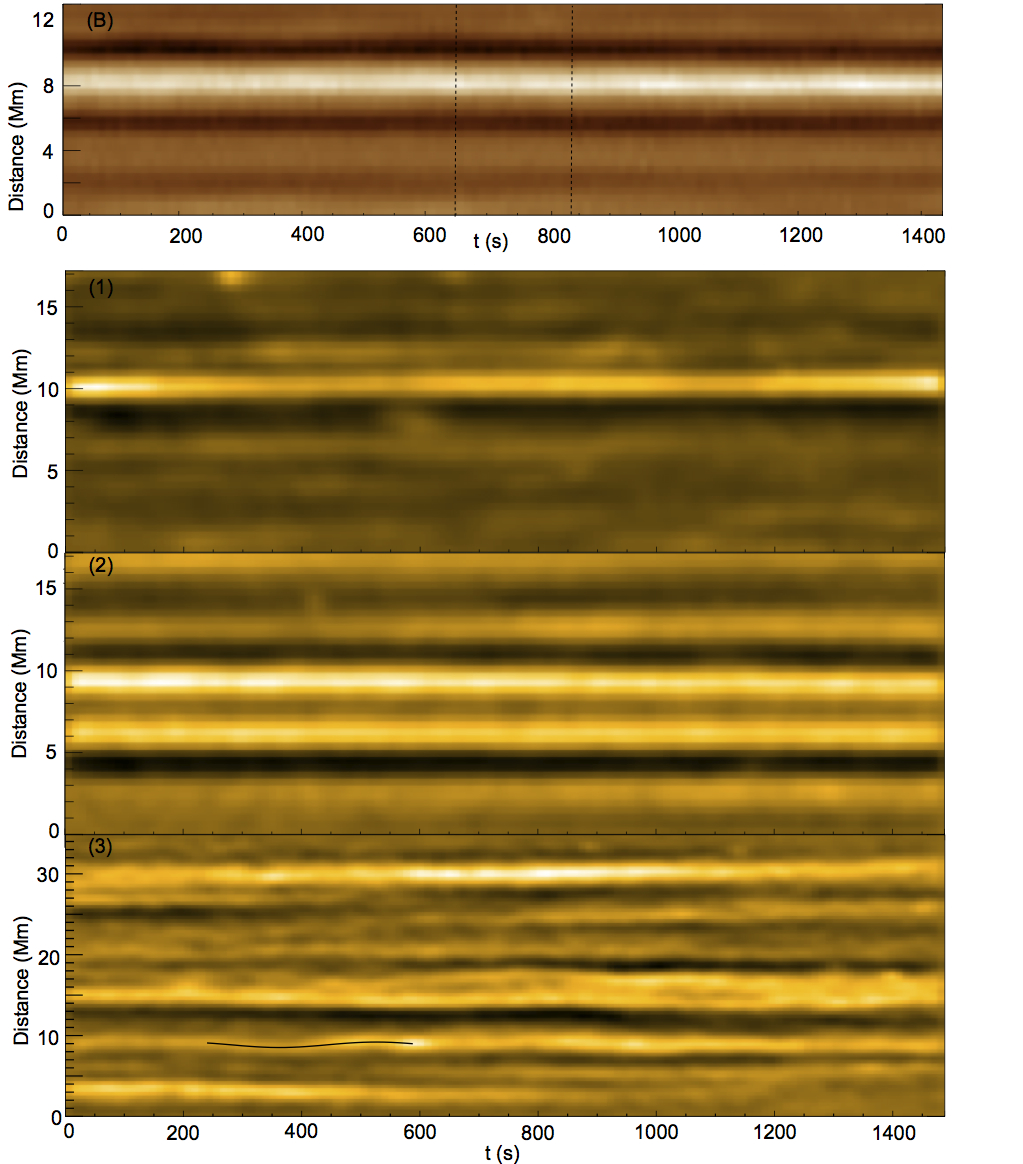} 
\caption{Time-distance diagram of the coronal feature labelled B as seen with AIA 193~{\AA} (\textit{top panel}). The time of the Hi-C observations are marked by the dashed lines. The second set of panels display time-distance diagrams for the coronal features identified as 1-3 in Fig.~\ref{fig:fov} as seen with AIA 171~{\AA}. The time-scales are given in seconds from the start of the AIA observations. 
}\label{fig:loop2_pt2_comb}
\end{figure}

In the second coronal structure labelled B (Fig.~\ref{fig:loop} - bottom row) there is 
no measurable periodic motion. {The results from the fitting hint at very small amplitude periodic motion 
(Fig.~\ref{fig:loop}: from 40-90~s), however the total displacement is below the magnitude of the 
uncertainties so cannot be claimed as wave motion.} The linear term in the sinusoidal fit, $g(t)$, may represent 
a transverse motion of the loop with a long period ($>230$~s). The apparent velocity amplitude of this transverse 
motion is $\sim$1~km\,s$^{-1}$, however, we do not claim it is periodic wave motion. In 
Fig.~\ref{fig:loop2_pt2_comb}, the time-distance diagram is shown for the AIA 193~{\AA}  data of the {same 
spatial location}. It demonstrates the activity for an extended period of time both before and after the Hi-C 
observations. Measurements suggest the structure does not exhibit large amplitude wave activity during the AIA 
observations (Fig.~\ref{fig:loop2_pt2_comb}). {However, we note that the coronal loop resolved in Hi-C has 
a width of $\sim230$~km. This is smaller than an AIA pixel and thus AIA does not resolve the structure 
adequately in Fig.~\ref{fig:loop2_pt2_comb}, which shows enhanced emission with a width of $\sim600$~km, 
possibly indicating contributions from neighbouring features.} 

{To gain a sense of typical wave activity in the active region}, additional coronal structures in the AIA $171$~{\AA} 
images are studied (labelled 1-3 in Fig.~\ref{fig:fov}). The features are partially visible in the Hi-C data only for a 
few frames due to the shift in pointing of the telescope. Fig.~\ref{fig:loop2_pt2_comb} displays the time-distance 
diagrams produced for these features. The cross-cuts from structures 1 and 2 do not show any visible signs of 
transverse displacement. {Applying the fitting routine suggests the presence of long period ($P>300$~s) 
and small amplitude ($A<100$~km, $v<3$~\kms) waves.} However, the feature labelled 3 is different and 
displays 
transverse motion that is visible even to the eye (online movie 5). A cross-cut is placed near the apex of this 
feature (close to arrow in Fig.~\ref{fig:fov}) and reveals periodic motion in numerous loop threads. One example is 
highlighted (over plotted black line) and the measured values are P=324$\pm$2~s, A=331$\pm$4~km, 
$v=6.42\pm0.09$~km/s. However, the transverse wave motion does not appear to be continuous and only one 
or two wave periods are visible before the perturbation decays. Note that this decay may not be due to damping, 
but rather that the observed motion is a wave-packet of finite length, which travels through the cross-cut.

\section{DISCUSSION AND CONCLUSIONS}
{With the five fold increase in resolution of Hi-C we have been able to observe transverse periodic motion 
($A<50$~km) in coronal loops 
that would otherwise have been difficult with AIA. The Hi-C coronal loops have widths on the order of or below AIA resolution and they are not resolved by AIA. The Hi-C data only reveals small amplitude, low energy waves and some coronal structures do not show measurable periodic transverse motion even at high resolution. }

The lack of identifiable wave motion in some Hi-C coronal loops could be limited by resolution still or 
data noise. Let us assume that the corona is filled with transverse waves and wave energy. Our results then allow 
us to place an upper bound on the amplitude of the transverse waves in some coronal active region structures. We 
suggest that transverse waves with an amplitude of $<0''.03\,(20~\mbox{km})$ (i.e., total displacement is less 
than half a pixel $\sim40$~km) are at the edge of Hi-C observational limitations. Hence, any low frequency waves 
present with periods in the range $50-200$~s will have velocity amplitudes of $<3$~km\,s$^{-1}$.

A further reason for the absence of periodic transverse waves in some loops is due to the short length of the data 
set, {$\approx$170~s (excluding the first seven frames). We assume that a wave 
can be detected (and fitted) if we can observe $3/4$ of a period. Hence, sinusoidal motions with periods 
$\gtrsim230$~s are unable to be detected here.} Signatures of long period transverse waves could be 
detected through the non-periodic component of the fit, $g(t)$.  The transverse displacement 
observed in this manner also appear to have relatively small velocity amplitudes. 
Further, waves with periods between 200-500~s and amplitudes >3~\kms would displace the loops central 
axis by at least 100-400~km, something which would be measurable in both Hi-C and AIA, but is not seen. 

By 
comparing an extended time-series taken from AIA with the short Hi-C time-series (e.g., 
Fig.~\ref{fig:loop2_pt2_comb}), it suggests that the period of time observed by Hi-C is representative of the 
dynamics in the structures studied, i.e., wave activity is low for extended periods of time. Small amplitude wave 
activity is also observed in coronal structures identified only in AIA (1 and 2  in Fig.~\ref{fig:fov}). Thus, 
our results suggest transverse waves supported by some active region coronal structure may have small velocity 
amplitudes ($<3$~\kms) and would not provide a significant heating contribution.

Previous observational evidence for transverse waves in active regions provided by measuring Doppler oscillations 
with Hinode EIS (\citealp{ERDTAR2008}, \citealp{VANetal2008c}, \citealp{TIAetal2012}), reporting velocity 
amplitudes 1-2~km\,s$^{-1}$ and periods of $\sim$300~s. If these previous observations are 
indeed signatures of transverse waves then they are consistent with our Hi-C and AIA amplitude 
measurements and our upper bound. These values are also consistent with CoMP results if we assume that a 
significant amount of wave energy ($>90\%$) is unresolved in the CoMP data. As with previous studies, we are 
limited to the observation of a few active region structures for a relatively short 
period of time. Our results hint that there may be categories of propagating transverse waves. Namely, waves with larger velocity amplitudes 
($v>$3~km\,s$^{-1}$) that are potentially excited by infrequent events which provide more energy flux (e.g., feature 3) than the driver of continuous (as suggested by CoMP results), small amplitude waves (<3~\kms). This idea is 
supported by our observations, with some structures demonstrating small amplitude periodic motion (e.g., 
features A, B, 1, 2) and others displaying one to two periods of large amplitude motion before the displacement 
becomes undetectable (feature 3). However, extended studies are required to confirm this.

\begin{acknowledgements}
RM is grateful to Northumbria University for the award of the Anniversary Fellowship and thanks A. Winebarger for 
useful discussions. The authors acknowledge IDL support provided by STFC. We 
acknowledge the High resolution Coronal imager instrument team for making the flight data publicly available. 
MSFC/NASA led the mission and partners include the Smithsonian Astrophysical Observatory in Cambridge, Mass; 
Lockheed Martin's Solar Astrophysical Laboratory in Palo Alto, Calif; the University of Central Lancashire in 
Lancashire, UK; and the Lebedev Physical Institute of the Russian Academy of Sciences in Moscow.
\end{acknowledgements}

\bibliographystyle{aa}

\end{document}